\documentclass[doublecol]{epl2}
\usepackage[english]{babel}
\usepackage{epsfig}
\usepackage{amsmath}
\usepackage{comment}
\usepackage{graphicx}
\usepackage{nicefrac}

\newcommand{\bs}{\boldsymbol}

\title{Theory for the single-point velocity statistics of fully developed turbulence}

\author{Michael Wilczek\inst{1}\thanks{E-mail: \email{mwilczek@uni-muenster.de}} \and Anton Daitche\inst{1} \and Rudolf Friedrich\inst{1}}
\shortauthor{M. Wilczek \etal}

\institute{\inst{1} Institute for Theoretical Physics, Westf\"alische Wilhelms-Universit\"at,\\ Wilhelm-Klemm-Str. 9, 48149 M\"unster, Germany}


\date{\today}

\abstract{
We investigate the single-point velocity probability density function (PDF) in three-dimensional fully developed homogeneous isotropic turbulence within the framework of PDF equations focussing on deviations from Gaussianity. A joint analytical and numerical analysis shows that these deviations may be quantified studying correlations of dynamical quantities like pressure gradient, external forcing and energy dissipation with the velocity. A stationary solution for the PDF equation in terms of these quantities is presented, and the theory is validated with the help of direct numerical simulations indicating sub-Gaussian tails of the PDF.}

\begin{document}

\maketitle
Understanding the statistical properties of fully developed turbulence is one of the main challenges in classical non-equilibrium statistical mechanics \cite{cardy08book,falkovich06pht}. A major obstacle for the formulation of a successful theory starting from the basic equations of fluid mechanics is the generically non-normal statistics of physical observables like, e.g., velocity increments across scales. This non-normal statistic renders the applications of well-known theoretical tools like renormalized perturbation expansions and the application of the renormalization group theory, which have been highly successful in the field of equilibrium phase transitions and critical phenomena, rather difficult \cite{forster77pra}. Hence the development of a statistical theory based on first principles remains one of the most fundamental problems of the field.\par
In contrast to the complex intermittent behavior of the velocity increment statistics, for which various phenomenological theories have been proposed (see, e.g., \cite{frisch95book} for an overview), it is a well-known fact that the single-point velocity PDF in homogeneous isotropic turbulence is close to Gaussian \cite{batchelor53book}, a fact which is often regarded as rather trivial. While Gaussianity often is assumed on a phenomenological basis, careful numerical and experimental investigations, however, suggest slight deviations from Gaussianity \cite{vincent91jfm,noullez97jfm,gotoh02pof} with a tendency to sub-Gaussian tails. The question to be raised towards fundamental turbulence research consequently is, whether the PDF indeed deviates from normality and if so, why. A number of works aim at answering this question with a somewhat inconclusive outcome. Jimenez suggests sub-Gaussian tails taking into account the precise shape of the energy spectrum \cite{jimenez97jfm}, while the instanton formalism is used by Falkovich and Lebedev \cite{falkovich97prl} to argue in favor of sub-Gaussian tails depending on the external forcing. In contrast to that Gaussian PDFs have been obtained in the case of decaying turbulence by Ulinich and Lyubimov \cite{ulinich69spj} and later by Hosokawa \cite{hosokawa08pre} or in the case of the cross-independence hypothesis by Tatsumi and coworkers \cite{tatsumi04fdr}. The latter works make use of the statistical framework of the Lundgren-Monin-Novikov hierarchy \cite{lundgren67pof,novikov68sdp,monin67pmm} with additional closure approximations.\par
The scope of the present work is to give a comprehensive and clear answer to the above formulated question. Moreover, we shall demonstrate in the present Letter that already the nearly Gaussian behavior of the most fundamental PDF is a consequence of the interplay of several non-trivial correlations. To this end, we make use of the LMN hierarchy combined with conditional averaging deriving a PDF equation for the single-point velocity PDF. This allows to recast the closure problem of turbulence in terms of statistical correlations between the velocity and the various dynamical quantities like pressure gradient, external forcing and the rate of energy dissipation. While this type of equation has also a long tradition in the context of PDF modeling of reacting flows (see, e.g., \cite{pope81pof,pope94arf,minier97pof,pope00book}), a profound combined theoretical and numerical analysis is still lacking. This brings us to the structure of the present article. After introducing the PDF equation for homogeneous isotropic turbulence, we will take into account statistical symmetries to simplify its functional form. This will allow to derive an exact stationary solution, which expresses the velocity PDF in terms of the above correlations. It turns out that the unknown correlations, which show up in terms of conditional averages, are subject to a number of constraints, which motivate a Gaussian solution as a lowest-order approximation closely related to the assumption of statistical independence of the various quantities. These theoretical aspects then will be supplemented by results from direct numerical simulations (DNS). The joint investigation then will show that deviations from this lowest-order approximation occur and that a subtle interplay of correlations leads to a sub-Gaussian shape of the PDF.\par
We start from the incompressible Navier-Stokes equation
\begin{equation}\label{eq:navier-stokes}
  \frac{\partial}{\partial t}\bs u+\bs u\cdot\nabla \bs u=-\nabla p+\nu \Delta \bs u+\bs F .
\end{equation}
Here $\bs u(\bs x,t)$ denotes the velocity field, $p(\bs x,t)$ is the pressure, $\nu$ denotes the kinematic viscosity and $\bs F(\bs x,t)$ denotes a large-scale forcing applied to the fluid to produce a statistically stationary flow. PDF methods allow to derive evolution equations for turbulent quantities right from the basic equation of motion. It can be shown \cite{lundgren67pof,novikov68sdp,monin67pmm,pope00book,wilczek09pre} that the evolution equation for the single-point PDF $f(\bs v;t)=\big\langle \delta(\bs u(\bs x,t)-\bs v) \big\rangle$ of the velocity for homogeneous turbulence may be derived taking the form 
\begin{equation}\label{eq:pdfvelhomo2}
  \frac{\partial}{\partial t}f=-\frac{\partial}{\partial v_i} \bigg \langle -\frac{\partial}{\partial x_i} p +  F_i  \bigg | \bs v  \bigg \rangle f - \frac{\partial }{\partial v_i}\frac{\partial }{\partial v_j} \bigg \langle \nu \frac{\partial u_i}{\partial x_k} \frac{\partial u_j}{\partial x_k}  \bigg | \bs v  \bigg \rangle f ,
\end{equation}
where the right-hand side is governed by the conditional averages of the pressure gradient, the external forcing and the so-called conditional dissipation tensor with respect to the velocity vector. Here we have introduced the sample space variable $\bs v$ for the velocity field. In the following we will restrict ourselves to the special case of homogeneous isotropic turbulence.  In this case the PDF of the velocity vector is fully determined by the PDF of its magnitude, $\tilde f(v)$, by the simple relation
\begin{equation}\label{eq:isopdf}
  \tilde f(v) = 4 \pi v^2 f(\bs v),
\end{equation}
and eq. \eqref{eq:pdfvelhomo2} may be recast taking the simple form
\begin{equation}\label{eq:pdfvelhomoiso}
  \frac{\partial}{\partial t} \tilde f=-\frac{\partial}{\partial v} \left( \Pi+\Phi-\frac{2\mu}{v} \right) \tilde f - \frac{\partial^2}{\partial v^2} \lambda \tilde f .
\end{equation}
Here the pressure and forcing term are expressed as
\begin{eqnarray}
 \left\langle -\nabla p \big | \bs v \right\rangle &= \Pi(v) \hat{\bs v} 
 \qquad \Pi(v)&=\left\langle - \hat{\bs u} \cdot \nabla p \big | v \right\rangle \label{eq:iso-pressure}   \nonumber  \\
 \left\langle \bs F \big | \bs v \right\rangle &= \Phi(v) \hat{\bs v}
 \qquad \Phi(v)&=\left\langle \hat{\bs u} \cdot \bs F \big | v \right\rangle \label{eq:iso-force} ,
\end{eqnarray}
i.e., as the conditionally averaged projections on the direction of the velocity $\hat{\bs u}$. The conditional dissipation tensor is characterized by two scalar functions
\begin{eqnarray}
   \bigg \langle \nu \frac{\partial u_i}{\partial x_k} \frac{\partial u_j}{\partial x_k}  \bigg | \bs v  \bigg \rangle &=& \mu(v) \, \delta_{ij} + \big[ \lambda(v)-\mu(v)\big]  \frac{v_i v_j}{v^2} \label{eq:D-lambda-mu} \nonumber \\
 \mu(v) &=& \frac{1}{4}\left\langle  \varepsilon+\nu \omega^2 \big | v\right\rangle-\frac{1}{2}\left\langle \nu(\mathrm{A}^T\hat{\bs u})^2\big | v\right\rangle \label{eq:mu_relation} \nonumber \\
 \lambda(v) &=& \left\langle \nu(\mathrm{A}^T\hat{\bs u})^2\big | v\right\rangle\label{eq:lambda_relation} ,
\end{eqnarray}
which are related to the conditional average of the rate of energy dissipation, the enstrophy and the additional term $\left\langle \nu(\mathrm{A}^T\hat{\bs u})^2\big | v\right\rangle$ related to the stretching and turning of the velocity vector by the velocity gradients $A_{ij}=\frac{\partial u_i}{\partial x_j}$. Due to the statistical symmetries eq. \eqref{eq:pdfvelhomo2} represents an effectively one-dimensional problem with the appearing conditional averages depending only on the magnitude $v$ of the velocity vector, cf. eqs. \eqref{eq:pdfvelhomoiso}--\eqref{eq:lambda_relation}. The most interesting feature of eq. \eqref{eq:pdfvelhomoiso} now is that it possesses a unique stationary solution
\begin{equation}\label{eq:statsol}
  \tilde f(v)=\frac{{\cal N}}{\lambda(v)} \exp \int_{v_0}^v \mathrm{d}v' \, \frac{ -\Pi(v')-\Phi(v')+\frac{2}{v'}\mu(v')}{\lambda(v')} ,
\end{equation}
expressing the PDF of the magnitude of velocity in terms of the introduced conditional averages. Due to eq. \eqref{eq:isopdf} this already quantifies the PDF of the full vector. The statement of this exact result is that the explicit shape of the PDF depends on the correlations of the different dynamical quantities with the velocity on the single-point level. Of course, this result still contains unclosed terms; its value lies in the fact that the quantities determining the shape of the stationary PDF are identified and can now be discussed on the basis of physical arguments. Alternatively, the conditional averages can be expressed in terms of the two-point velocity PDF, such that the problem can also be formulated in terms of multi-point PDFs. This, for example, has been done in \cite{lundgren72lnp,tatsumi04fdr,hosokawa08pre}. We here rather follow a discussion in terms of the conditional averages. The precise knowledge of these functions now would allow to determine the exact shape of the velocity PDF. However, they cannot be calculated explicitly without further closure assumptions, but numerical simulations or experiments may be used to obtain further insights. Nevertheless, a number of constraints can be formulated for these unknown functions as upon integration the conditional averages have to reduce to ordinary averages,
\begin{eqnarray}
   -\int_{0}^{\infty} \mathrm{d}v \, v\,\Pi(v) \tilde f(v) &=& \left \langle \bs{u}\cdot\nabla p \right \rangle = 0
  \label{pressure-int-constraint} \\
  \int_{0}^{\infty} \mathrm{d}v \, v\,\Phi(v) \tilde f(v) &=& \left\langle \bs u \cdot \bs F\right\rangle = \left\langle \varepsilon \right\rangle
  \label{Phi-int-constraint} \\
  \int_{0}^{\infty} \mathrm{d}v \, [\lambda(v)+2\mu(v)] \tilde f(v) &=& \left\langle \nu \frac{\partial u_i}{\partial x_k} \frac{\partial u_i}{\partial x_k} \right\rangle = \left\langle \varepsilon \right\rangle .\label{lambda-mu-int-constraint}
\end{eqnarray}
It further can be shown on the basis of isotropy that $\Pi(0)=\Phi(0)=0$, $\lambda(0)=\mu(0)$ and $\frac{\mathrm{d} \lambda}{\mathrm{d} v}\big|_{v=0}=\frac{\mathrm{d} \mu}{\mathrm{d} v}\big|_{v=0}$ = 0, which further constrains the possible functional forms of the unclosed terms.\par
The closure problem now prohibits an exact analytical treatment without further assumptions. Before we estimate the unclosed terms from our DNS data, we would like to demonstrate that Gaussian statistics can be obtained as the result of a number of simple modeling assumptions. The purpose of this short presentation, however, is not to claim the velocity statistics to be exactly Gaussian. It rather shows that physically intuitive, but in the end inappropriate, assumptions can lead to quite reasonable results.\par
To proceed with an analytical treatment of the PDF equation, one now, on the one hand, could assume decoupling of the velocity field that varies on large scales and the dissipation field, which varies on much shorter scales. As a consequence, the conditional dissipation tensor is a constant isotropic tensor. The pressure gradient cannot be treated with such a simple argument, because it is known to be a strongly fluctuating, but long-range correlated quantity. Still, to comply with the integral constraint \eqref{pressure-int-constraint}, the most simple modeling assumption is a vanishing contribution, which is related to the assumption of statistical independence of the pressure gradient and velocity field. Finally, the external forcing and the velocity field both vary on large scales, such that mutual dependences have to be expected. These considerations lead to a lowest-order approximation (in terms of powers of $v$) compliant with the above constraints, taking the form
\begin{eqnarray}
  \Pi_0(v) &=& 0 \label{eq:gaussianclosure1} \\
  \Phi_0(v) &=& \frac{\langle \varepsilon \rangle}{3\sigma^2}v \label{eq:gaussianclosure2}  \\
  \lambda_0(v) = \mu_0(v)  &=& \frac{\langle \varepsilon \rangle}{3}, \label{eq:gaussianclosure3}
\end{eqnarray}
where $\sigma = u_{\mathrm{rms}} =\sqrt{\frac{\langle \bs u^2 \rangle}{3} }$ denotes the standard deviation of the velocity. That means, in this lowest-order approximation the conditional pressure gradient vanishes, and the external forcing depends linearly on the velocity on the statistical average. The terms related to the conditional dissipation tensor are constant, viz. statistically independent of the velocity field. Insertion into eq. \eqref{eq:statsol} yields the Gaussian solution
\begin{equation}\label{gaussian}
  f(\bs{v})=\frac{1}{(2\pi\sigma^{2})^{3/2}}\exp\left( -\frac{1}{2}\frac{\bs{v}^2}{\sigma^2} \right),
\end{equation}
i.e., under a lowest-order ansatz including the assumption of negligible correlations we have found a closure of eq. \eqref{eq:pdfvelhomo2} that predicts Gaussian velocity distributions.\footnote{In fact, any closure with $\Pi_0(v)+\Phi_0(v)=\frac{\langle \varepsilon \rangle}{3\sigma^2}v$ and $\lambda_0(v)=\mu_0(v)=\frac{\langle \varepsilon \rangle}{3}$ leads to a Gaussian velocity PDF. The present one is additionally consistent with the integral constraints \eqref{pressure-int-constraint}--\eqref{lambda-mu-int-constraint} and the corresponding energy budget equation for stationary homogeneous isotropic turbulence.} In turn, the existence of non-negligible correlations will generically cause deviations from Gaussianity, which will be discussed in the following. In fact, the conditional averages obtained from our numerical simulations will significantly deviate from the modeled ones above and together will lead to deviations from Gaussianity.\par

\begin{table}
\begin{center}
  \begin{tabular}{@{\quad}c@{\quad}c@{\quad}c@{\quad}c@{\quad}c@{\quad}c@{\quad}c@{\quad}c}
    \hline 
    $N^3$ & $R_{\lambda}$ & $L$ & $T$ & $u_{\mathrm{rms}}$ & $\nu$ & $\left\langle \varepsilon \right\rangle$ & $ k_{\mathrm{max}}\eta$\\
    \hline 
    $512^3$  &  112 & 1.55 & 2.86 & 0.543 & $10^{-3} $ & 0.103 & 2.03\\
    \hline
  \end{tabular}
\end{center}
\caption{Major simulation parameters. Number of grid points $N^3$, Reynolds number based on the Taylor micro-scale $R_{\lambda}$, integral length scale $L$, large-eddy turnover time $T$, root-mean-square velocity $u_{\mathrm{rms}}$, kinematic viscosity $\nu$, average rate of energy dissipation $\left\langle \varepsilon \right\rangle$. $ k_{\mathrm{max}}\eta$ characterizes the spatial resolution of the smallest scales.}
\label{tab:simpara}
\end{table}

\begin{figure}
\centering\includegraphics[width=0.48\textwidth]{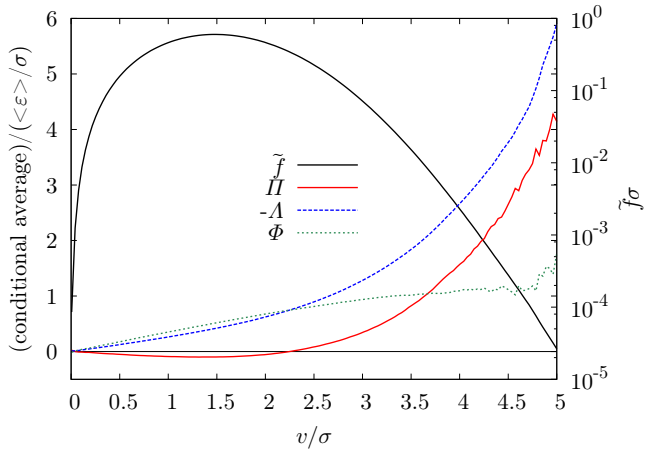}
\caption{The conditional averages $\Pi(v)$, $\Lambda(v)$, $\Phi(v)$ and the PDF $\tilde f(v)$ shown for reference. The conditional averages show an explicit $v$-dependence indicating statistical correlations.}
\label{fig:PILAPHI}
\end{figure}

\begin{figure}
  \centering\includegraphics[width=0.48\textwidth]{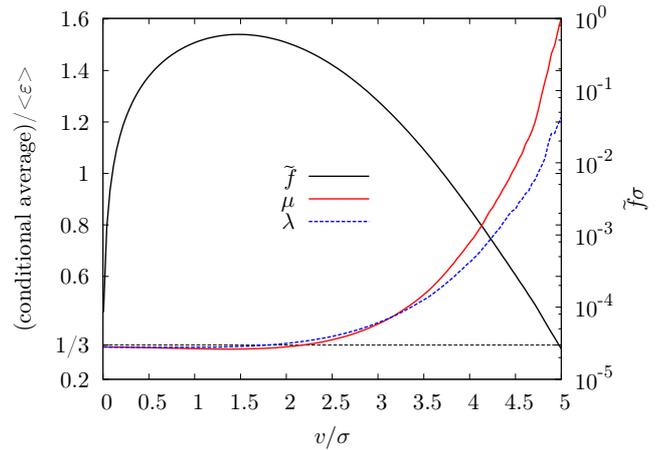}
  \caption{The conditional averages $\mu(v)$, $\lambda(v)$ and the PDF $\tilde f(v)$. The statistical correlations increase in the tail of the PDF.}
  \label{fig:mulaepsz}
\end{figure}

\begin{figure}
  \centering\includegraphics[width=0.48\textwidth]{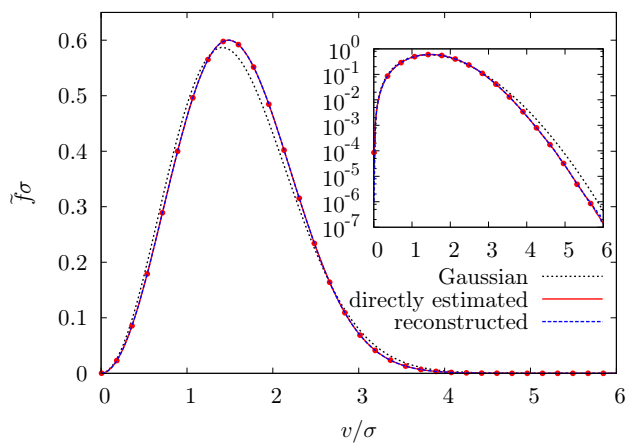}
  \caption{Comparison of the PDF $\tilde f(v)$ with the evaluation of eq. \eqref{eq:statsol}. The reconstructed PDF collapses with the directly estimated one. The excellent agreement demonstrates the consistency of the theoretical results. A comparison with an angle integrated Gaussian shows that, although being close to Gaussian, significant deviations exist.}
  \label{fig:recpdf}
\end{figure}

\begin{figure}
  \includegraphics[width=0.48\textwidth]{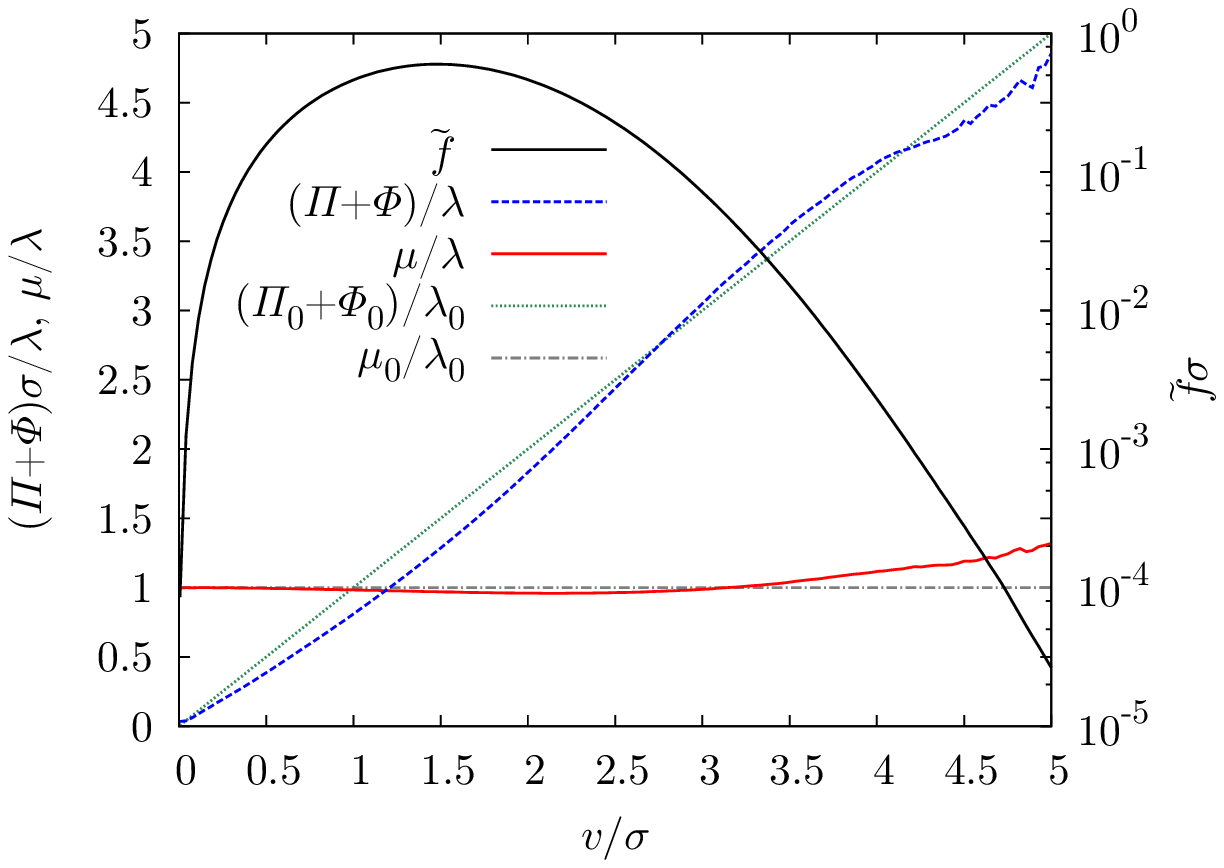}
  \caption{The quotients $(\Pi+\Phi)/\lambda$ and $\mu/\lambda$ appearing in eq. \eqref{eq:statsol} show an approximately linear and constant functional form, respectively. The corresponding quotients expected from the analytical Gaussian closure approximation \eqref{eq:gaussianclosure1}--\eqref{eq:gaussianclosure3} are shown for comparison.}
  \label{fig:integrand-analyzed}
\end{figure}

The numerical simulations are obtained with a standard pseudospectral code in a cubic box with periodic boundary conditions, for which the main simulation parameters are summed up in table \ref{tab:simpara}. The results were checked within a range of Reynolds numbers up to $R_{\lambda}\approx 220$ at a resolution of $1024^3$ and for several large-scale forcing methods. The results of the analysis proved to be robust in the sense that all of the presented results do not vary strongly within the range of investigated Reynolds number and the different types of forcing schemes. It would, however, be of general interest to obtain the statistical quantities presented in the following for flows at significantly higher Reynolds number in order to study Reynolds number effects. Due to its excellent statistical quality we present the results of a $512^3$ simulation, where the statistics has been evaluated for a run longer than $150$ large-eddy turnover times sampled sufficiently often (about 200 snapshots have been stored) to form a proper statistical ensemble. As derivatives of the velocity field like the energy dissipation have to be evaluated, special emphasis has been put on an appropriate resolution of the smallest scales of the flow. \par

In the following we will evaluate the terms that determine the shape of the velocity PDF according to eq. \eqref{eq:statsol}. The conditional pressure gradient term and the conditional forcing term are shown in fig. \ref{fig:PILAPHI}. The most striking feature is that the pressure contribution is non-vanishing. For low values it is negative indicating that fluid particles on average will be decelerated due to the (negative) pressure gradient. For high values of velocity, however, the term becomes strongly positive, which correspondingly indicates an acceleration of fluid particles with high velocities due to this term. If the pressure gradient term is non-vanishing, the observed zero-crossing is necessary to fulfill the integral constraint \eqref{pressure-int-constraint}. These findings show that pressure contributions may not be neglected as opposed to the analysis presented in \cite{tatsumi04fdr,hosokawa08pre}. The forcing term is positively correlated with the velocity and hence also has an accelerating effect on a fluid particle on the statistical average. For low values of velocity this effect is approximately linear and seems to saturate for higher velocities, which suggests that the forcing has a minor effect on the tails of the PDF. It can be shown in the framework of PDF equations that the conditional acceleration, i.e. the conditional right-hand side of the Navier-Stokes equation \eqref{eq:navier-stokes}, vanishes \cite{novikov93jfr}. For isotropic turbulence this implies
\begin{equation}
  \Pi(v)+\Phi(v)+\Lambda(v)=0,
\end{equation}
by which we have introduced the viscous forces $\Lambda(v)=\langle \nu \hat{\bs u} \cdot \Delta \bs u | v \rangle$. This term indeed has a decelerating effect on a fluid element as can be seen from  fig. \ref{fig:PILAPHI}. On the whole the pressure gradient and the external forcing are balanced by viscous forces.\par
For the stationary solution of the PDF equation we additionally have to take into account the terms from the conditional dissipation tensor. For PDF modeling and in the case of some closure theories \cite{ulinich69spj,pope81pof,tatsumi04fdr,minier97pof,hosokawa08pre} this tensor is taken to be a constant isotropic tensor of rank two, compliant with our lowest-order Gaussian approximation. Fig. \ref{fig:mulaepsz} shows the outcome of our DNS analysis. Indeed, we find the approximate relation $\lambda \approx \mu$ for the two eigenvalues holding especially well for moderate values of velocity indicating that this tensor is approximately isotropic. It is furthermore nearly constant for low values of velocity, such that the above approximation is appropriate for this region of the PDF. For high velocities, however, stronger statistical correlations are apparent showing that high velocities on average are correlated with dissipative events. Consequently, the tails of the PDF will be influenced by these correlations leading to a more rapid decay (recall, e.g., that $\lambda$ appears in eq. \eqref{eq:statsol} in the denominator). The fact that the conditional dissipation tensor is approximately isotropic indicates the absence of directional correlations of, e.g., the energy dissipation field with the velocity field. In view of the aforementioned results the assumption of a decoupling of large- and small-scale fields may be regarded as a rough approximation leading to a Gaussian closure, and one is likely to neglect important features of the statistics of turbulence under this assumption.\par
These results can be used to motivate improved closures by modeling the unclosed terms,  such that they resemble the functional shape observed in the DNS. This, of course, brings up the question if the observed results are of universal nature, and it would be interesting to perform the analysis also for different flows apart from the idealized situation of homogeneous isotropic turbulence. In the sense that we expect statistical correlations between, e.g., the dissipation and velocity field also in more complex situations, we think that the obtained insights are of more general nature.\par
Now having obtained all the unclosed terms from the DNS, we can explicitly evaluate eq. \eqref{eq:statsol}, which is presented in fig. \ref{fig:recpdf}. As all of the presented results leading to eq. \eqref{eq:statsol} are exact, the PDF calculated from eq. \eqref{eq:statsol} perfectly collapses with the velocity PDF directly determined from the DNS data. This consistency check proves that the interplay of statistical correlations of the velocity with the dynamical effects of the pressure gradient, external forcing and dissipative terms causes the deviations from Gaussianity and especially the sub-Gaussian tails of the PDF. In view of the pronounced statistical correlations for large values of velocity it is actually surprising that deviations from Gaussianity are comparably small, which shall be explained in the following. A closer look at eq. \eqref{eq:statsol} shows that deviations from Gaussianity may arise for different reasons. Considering the exponential factor, deviations from Gaussianity appear if $\frac{\lambda}{\mu}$ is non-constant and $\frac{\Pi+\Phi}{\lambda}$ is non-linear. These two quantities are presented in fig. \ref{fig:integrand-analyzed} along with the corresponding quantities from the analytical Gaussian closure approximation \eqref{eq:gaussianclosure1}--\eqref{eq:gaussianclosure3}. It turns out that the strong correlations observed in figs. \ref{fig:PILAPHI} and \ref{fig:mulaepsz} tend to cancel to a certain extent, such that  the quotient of the sum of the (negative) pressure gradient and the external forcing over the eigenvalue $\lambda$ of the conditional dissipation tensor is approximately linear and $\frac{\mu}{\lambda}$ is approximately constant. The former is possible since regions of high velocities are on average associated both with large accelerating forces due to the pressure gradient and strong dissipative events; in these regions the forcing plays a minor role compared to the other dynamical influences. The latter approximate behavior comes due the fact that only weak directional correlations between the velocity field and the dissipation field occur. The slight deviations from linearity and constancy, respectively, of these quantities with their particular functional form eventually contribute to deviations from Gaussianity. Additional deviations arise due to the functional shape of $\lambda$, which appears in the prefactor of eq. \eqref{eq:statsol}, as the Gaussian closure approximation \eqref{eq:gaussianclosure3} here suggests a constant function. By and large the subtle interplay of the pronounced observed statistical correlations causes rather moderate but still significant deviations from Gaussianity towards sub-Gaussian tails.\par
To conclude, we have analyzed the single-point statistics of fully developed turbulence in the framework of the LMN hierarchy. Taking into account statistical symmetries, we have derived an expression for the stationary PDF, which highlights the connection between the shape of the PDF and correlations of the velocity with dynamical effects. This stationary solution is an exact result derived directly from the Navier-Stokes equation. One of the strengths of this approach is that the closure problem of turbulence is expressed in terms of correlations on the single-point level rather than the traditional approach in terms of a coupling to higher-order moments or multi-point PDFs, allowing for a physical discussion of the unclosed terms. Furthermore, we have found Gaussian solutions applying simple physical closure assumptions consistent with the constraints on the conditional averages.\par
To explain the observed deviations from Gaussianity, we have used DNS data to study the conditional averages that contribute to the shape of the PDF finding non-negligible correlations. Evaluation of the stationary solution of the PDF equation proved to be in perfect agreement with the directly estimated DNS data. The results demonstrate how a combined theoretical and numerical investigation can be used to give insights into the details of the statistics of turbulent flows. Apart from giving a comprehensive picture of the single-point velocity statistics, the results may motivate improved closure assumptions for PDF modeling by taking into account the functional shape of the conditional averages observed in DNS. It remains an interesting question for further research if the observed correlations carry over to more complex flow situations like, e.g., wall-bounded turbulence.\par
The theory can also be applied to decaying turbulence. An extended analysis including more details on the theoretical derivation, application to decaying turbulence and a comparison to the vorticity statistics will be the topic of a detailed report published elsewhere \cite{wilczek10sub}.

\acknowledgements{We would like to thank one of the referees for an interesting discussion and constructive input. Computational resources for this work were granted by the LRZ Munich, project No. h0963.}


\end{document}